\documentclass[preprint,showpacs,preprintnumbers,amsmath,amssymb]{revtex4}
\usepackage{graphicx}
\usepackage{dcolumn}
\usepackage{bm}

\begin{document}

\preprint{APS/123-QED}

\title{Pressure induced superconductor-insulator transition \\in the spinel compound CuRh$_{2}$S$_{4}$}

\author{M. Ito}
\email{showa@hiroshima-u.ac.jp}
\author{J. Hori}
\author{H. Kurisaki}
\author{H. Okada}
\author{A. J. Perez Kuroki}
\author{N. Ogita}
\author{M. Udagawa}
\author{H. Fujii}
\author{F. Nakamura}
\author{T. Fujita}
\author{T. Suzuki}%
\email{tsuzuki@hiroshima-u.ac.jp}
\affiliation{
Department of Quantum Matter, ADSM, Hiroshima University, Higashi-Hiroshima 739-8530, Japan.
}%

\date{\today}

\begin{abstract}
We performed resistivity measurements in CuRh$_{2}$S$_{4}$ under quasi-hydrostatic pressure of up to 8.0 GPa, and found a pressure induced superconductor-insulator (SI) transition.
Initially, with increasing pressure, the superconducting transition temperature $T_c$ increases from 4.7 K at ambient pressure to 6.4 K at 4.0 GPa, but decreases at higher pressures.
With further compression, superconductivity in CuRh$_{2}$S$_{4}$ disappears abruptly at a critical pressure $P_{\rm SI}$ between 5.0 and 5.6 GPa, when it becomes an insulator. 
\end{abstract}

\maketitle


 Unusual physical properties of chalcogenide-spinels have attracted current interest because of a new type of metal-insulator (MI) transition~\cite{rf:Radaelli} found in the thiospinel CuIr$_{2}$S$_{4}$, with a lattice parameter $a$ = 9.847 $\rm \AA$  at room temperature~\cite{rf:Furubayashi}.
 As far as the lattice parameter~\cite{rf:Hagino} is concerned, isomorphic CuRh$_{2}$S$_{4}$ with $a$ = 9.787 $\rm \AA$ and CuRh$_{2}$Se$_{4}$ with $a$ = 10.269 $\rm \AA$ are regarded as compressed and expanded respectively, compared to CuIr$_{2}$S$_{4}$.
 Both compounds, however, are well known as superconductors~\cite{rf:Hagino,rf:Van Maaren,rf:Shelton,rf:Bitoh,rf:Shirane} without an MI transition.
In spite of the similar temperature dependence of electric resistivity in the normal conducting state, the absolute value of the resistivity of CuRh$_{2}$S$_{4}$, with a smaller lattice parameter is about 20 times larger  than for CuRh$_{2}$Se$_{4}$, with a larger lattice parameter. 
This suggests that some anomaly may occur in the transport properties of CuRh$_{2}$S$_{4}$ as it is further compressed.
In this letter we report that CuRh$_{2}$S$_{4}$ makes a sudden transition from a superconductor to an insulator at a critical pressure $P_{\rm SI}$ between 5.0 and 5.6 GPa. We are not aware of a pressure induced superconductor-insulator transition having been previously observed in other materials.

 Polycrystalline CuRh$_{2}$S$_{4}$ was prepared by a direct solid-state reaction. Fine powders of Cu ( 99.999\% ), Rh ( 99.9\% ) and S ( 99.9999\% ) were mixed in stoichiometric ratio and were reacted in a sealed quartz tube at 850 $^{\circ}$C for 10 days.
After being pulverized, the specimen was pressed into pellets and sintered at 1000 $^{\circ}$C for 3 days.
Sample manipulation was always carried out in a glove box filled with purified argon to minimize oxidation of the pellets. 
 Electric resistivity $\rho(T)$  was measured by a standard four-probe method with increasing temperature $T$ from 4.2 to 300 K under quasi-hydrostatic pressure $P$ up to 8 GPa. 
 The pressure was applied by using a cubic-anvil device~\cite{rf:Mori} cooled in a liquid $^4$He cryostat.
To examine whether a pressure-induced structural transition occurs, Raman spectra under quasi-hydrostatic pressure up to 11.7 GPa were measured by a micro-Raman spectrometer with a diamond-anvil cell. An Ar laser with the wave length of 514.5 nm was used as the incident beam. A mixture of methanol and ethanol was used as the pressure-transmitting medium. The pressure inside the cell was determined from the shift of the fluorescence line of ruby.  
\begin{figure}
\includegraphics[height=142mm,clip]{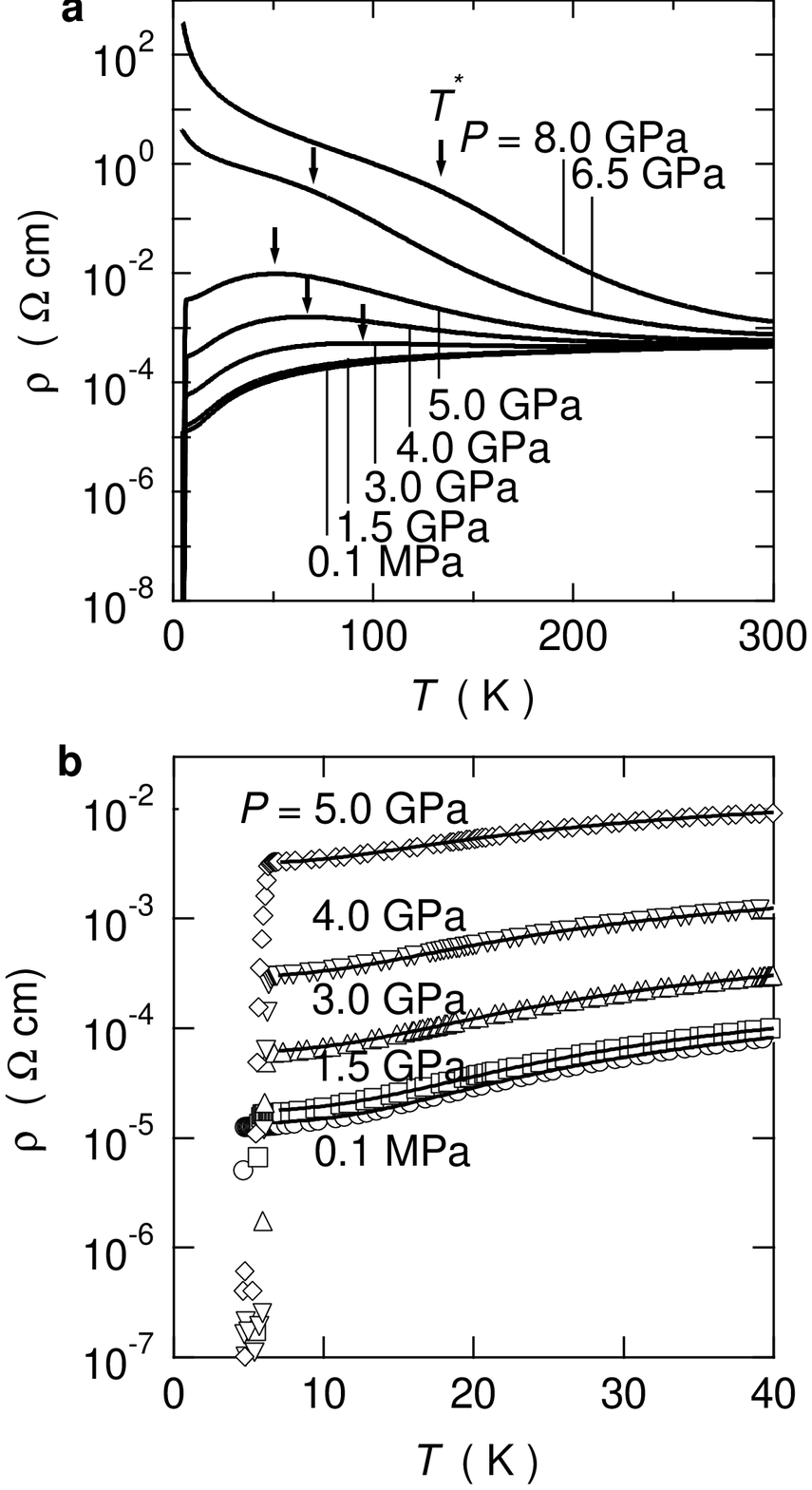}
\caption{\label{fig:fig.1} Temperature dependence of electric resistivity $\rho(T)$ under pressure up to 8.0 GPa. \textbf{a}, Resistivity in the temperature range from 4.2 to 300 K. The pressure was increased in the sequence of 0.1 MPa, 1.5, 3.0, 4.0, 5.0, 6.5 to 8.0 GPa. We re-measured $\rho(T)$ at 1.5 GPa after pressure was decreased from 8.0 GPa to confirm the reversibility with respect to the variation of pressure. The characteristic temperature $T^{\ast}$ is indicated by the arrows. Superconductivity is observed at $P \le $ 5.0 GPa. \textbf{b}, An expanded plot of  $\rho(T)$ in the range 4.2 $< T <$ 40 K at $P \le $ 5.0 GPa. The solid lines are the best fits with the phenomenological equation (1) using the parameters listed in Table 1.}
\end{figure}

Figure 1 shows the $T$-dependence of $\rho$ between 4.2 and 300 K at various pressures. The response of resistivity to pressure is quite unusual. 
At ambient pressure, $\rho(T)$ shows metallic $T$-dependence with $\partial \rho(T)/\partial T >$ 0 above the superconducting transition temperature $T_{\rm c}$ = 4.7 K defined by the onset of the resistance drop. 
As was pointed out by Hagino \textit{et al}.~\cite{rf:Hagino}, $\rho(T)$ in the normal state of CuRh$_{2}$S$_{4}$ is similar to that for A-15 compounds. The $T$-dependence can be fitted by a phenomenological equation~\cite{rf:Woodard,rf:Webb}:
\begin{equation}
 \rho(T) = \rho_{0} + \rho_{1} T + \rho_{2} \exp (-T_{0} / T),    
\end{equation}
where, $\rho_{0}$, $\rho_{1}$, $\rho_{2}$ and $T_{0}$ are $T$-independent fitting parameters.
The variation of $\rho(T)$ is monotonic under low pressure.
However as $P$ increases above 3.0 GPa, a broad peak appears in $\rho(T)$ at around a characteristic temperature $T^{\ast}$, which is indicated by arrows in Fig. 1. 
$T^{\ast}$ initially decreases with increasing $P$ and then starts to increase above 5.0 GPa. 
In the range 3.0 $\le  P \le$  5.0 GPa, $\partial \rho(T)/\partial T$ changes its sign from negative to positive at $T^{\ast}$ as $T$ decreases. 
$\rho(T)$ shows metallic $T$-dependence ($\partial \rho(T)/\partial T >$ 0) at low temperatures, with the superconducting transition occurring above 4.2 K. 
The metallic variation in the range $T_{\rm c} < T < T^{\ast}$ can be fitted by the equation (1) as shown by the solid lines in Fig. 1b, with the parameters listed in Table 1. 
Application of pressure enhances $\rho_{0}$ drastically to a 250 times larger value. 
A more important point is that superconductivity completely disappears for $P \ge$ 6.5 GPa, where $\rho(T)$ shows non-metallic $T$-dependence ($\partial \rho(T)/\partial T <$ 0) over the whole temperature range measured. 
As shown in Fig. 2, it is also remarkable that $\rho(10\rm\ K)$ is greatly enhanced, by over 7 orders of magnitude, with increasing pressure from 0.1 MPa to 8.0 GPa. 
This enhancement is highly anomalous. Compression usually gives rise to a reduction in resistivity in many materials and in some even to superconductivity~\cite{rf:Shimizu}, because overlapping of electronic wave functions among neighboring atoms is promoted by compression. 
Well-known examples of the pressure effect due to this electronic origin are found in the Mott transition~\cite{rf:Nakamura} or Wilson transition~\cite{rf:Blachan,rf:Riggleman}. 
Even in random systems, a transition from an insulator to a metal was found by applying  uniaxial compressive stress~\cite{rf:Paalanen}.
\begin{table*}
\caption{\label{tab:table1}Fitting parameters $\rho_0$, $\rho_1$, $\rho_2$ and $T_0 $ evaluated for the low temperature part of $\rho(T)$ under pressures from 0.1 MPa to 5.0 GPa.}
\begin{ruledtabular}
\begin{tabular}{cccccc}
  $P$ ( GPa )     &    $\rho_0$ ( $\Omega $cm )  &    $\rho_1$ ( $\Omega $cm/K )  &    $\rho_2$  ( $\Omega $cm ) &    $T_0 $ (K) \\
\hline
$\sim$10$^{-4}$ & 1.214$\times 10^{-5}$ & 2.420$\times 10^{-7}$ & 3.140$\times 10^{-4}$ & 65.9 \\  
 1.5             & 1.512$\times 10^{-5}$ & 3.913$\times 10^{-7}$ & 3.567$\times 10^{-4}$ & 65.8	\\ 
 3.0             & 5.341$\times 10^{-5}$ & 1.256$\times 10^{-6}$ & 9.000$\times 10^{-4}$ & 60.4	\\ 
 4.0             & 2.682$\times 10^{-4}$ & 5.048$\times 10^{-6}$ & 2.912$\times 10^{-3}$ & 53.4	\\
 5.0             & 3.047$\times 10^{-3}$ & 2.374$\times 10^{-5}$  & 1.561$\times 10^{-2}$ & 43.1 \\ 
\end{tabular}
\end{ruledtabular}
\end{table*}
\begin{figure}
\includegraphics[height=72mm,clip]{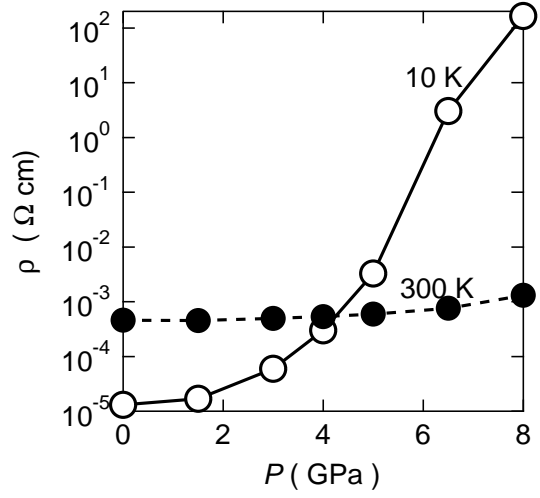}
\caption{\label{fig:fig.2} Pressure dependence of electric resistivity $\rho(T)$. The open and closed circles represent $\rho(10\rm\ K)$ and $\rho(300\rm\ K)$, respectively. 
Each point was taken from the experimental data shown in Fig. 1(a). Both $\rho(10\rm\ K)$ and $\rho(300\rm\ K)$ basically increase monotonically with pressure, although an inflection point is seen in $\rho(10\rm\ K)$ between 5.0 and 6.5 GPa. }
\end{figure}
\begin{figure}
\includegraphics[height=72mm,clip]{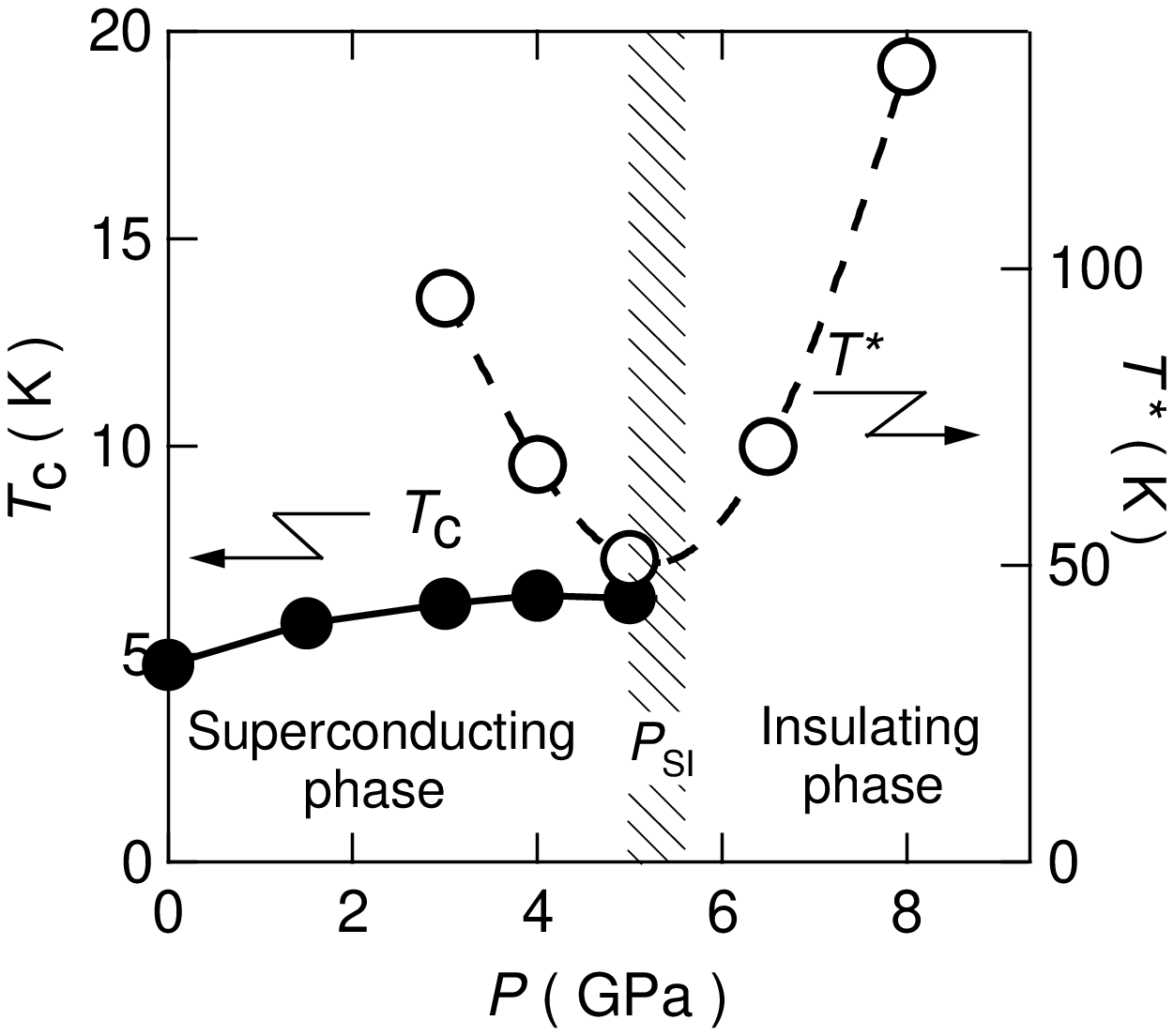}
\caption{\label{fig:fig.3} The superconducting transition temperature $T_{\rm c}$ and the characteristic temperature $T^{\ast}$ plotted as functions of pressure. The solid and broken lines are guides to the eyes. Superconductivity vanishes at $P_{\rm SI}$, where $T^{\ast}$ has the minimum value and the metal-insulator phase boundary (hatched) is expected to lie.}
\end{figure}
The effects of compression on $T_{\rm c}$ and $T^{\ast}$ are summarized in Fig. 3. 
The $T_{\rm c}$ value increases with increasing $P$, which is consistent with the previous work~\cite{rf:Shelton} carried out under pressures lower than 2.0 GPa. 
 With further compression, $T_{\rm c}$ starts to decrease after having a maximum value of 6.4 K at 4.0 GPa. 
 Around 5.3 GPa, where $T^{\ast}$ has a minimum value, the low-temperature transport properties of CuRh$_{2}$S$_{4}$ change from superconducting to insulating.
 We believe that this SI transition is a bulk property of CuRh$_{2}$S$_{4}$.
 The slight depression of $T_{\rm c}$ observed above 4.0 GPa might be due to a precursor or fluctuation of the SI transition.
 To our knowledge, this is the first report on a pressure induced SI transition.
 
The $P$-variation of $\rho(T)$ over the whole temperature range shown in Fig. 1 further suggests that the SI transition occurs at the same $P$ that the sample changes from a metal to an insulator. The SI transition presumably results from the disappearance of carriers at the MI transition between 5.0 and 5.6 GPa. 
The mechanism of the MI transition is still unclear in the present case, and one may consider other origins for the transition besides the localization of carriers~\cite{rf:Liu}.

Here, let us discuss the pressure induced MI transition from the viewpoint of change in the band structure both with and without change in the crystalline lattice. 
In the first case, one can consider a transition driven by charge ordering, as for the new type MI transition reported in CuIr$_{2}$S$_{4}$ at $T_{\rm MI}$ = 226 K~\cite{rf:Radaelli}.
As mentioned above, CuRh$_{2}$S$_{4}$ is a compressed version of CuIr$_{2}$S$_{4}$ and therefore has a similar electronic structure to CuIr$_{2}$S$_{4}$.
The Rh ions in CuRh$_{2}$S$_{4}$ occupy the octahedral sites and have valences Rh$^{3+}$ (electronic configuration of 4$d\epsilon^{6}d\gamma ^{0}$ ) and Rh$^{4+}$ (4$d\epsilon^{5}d\gamma ^{0}$ )~\cite{rf:Hagino}, analogous to Ir$^{3+}$ (5$d\epsilon^{6}d\gamma ^{0}$ ) and Ir$^{4+}$ (5$d\epsilon^{5}d\gamma ^{0}$ ) in CuIr$_{2}$S$_{4}$~\cite{rf:Furubayashi}. 
As the temperature decreases, CuIr$_{2}$S$_{4}$ undergoes a transition from a metal to an insulator at $T_{\rm MI}$, which is accompanied by a structural change from cubic to tetragonal with a volume contraction~\cite{rf:Radaelli,rf:Furubayashi} of 0.7\%. In this case, compression stabilizes the insulating phase with charge ordering as was evidenced by the elevation of $T_{\rm MI}$ with increasing pressure~\cite{rf:Oomi}. 
The absolute value of resistivity above $T_{\rm MI}$ is also enhanced substantially by this charge ordering.
If we assume the same mechanism, we may easily predict that as a consequence of the charge ordering of Rh$^{3+}$ and Rh$^{4+}$, CuRh$_{2}$S$_{4}$ undergoes an MI transition above 300 K by compression, and a discontinuity should be detected in the $P$-dependence of $\rho(300\ \rm K)$. 
However, no sign of such a jump is found in $\rho(300\ \rm K)$ shown in Fig. 2. 
Consequently, the MI transition mechanism of CuIr$_{2}$S$_{4}$ type is ruled out for CuRh$_{2}$S$_{4}$.
To confirm the discussion, we performed the structure-sensitive micro-Raman spectroscopy under pressure in the energy range between 200 and 600 cm$^{-1}$ at 300 K. The laser beam was focused on one of the small single-crystals in the polycrystalline sample.
Figure 4 shows Raman spectra for selected pressures. For a spinel structure, five phonons (A$_{1g}$, E$_{g}$ and three T$_{2g}$) are Raman active~\cite{rf:Bruesch}. One of the T$_{2g}$ phonons usually has the lowest energy with around 100 cm$^{-1}$~\cite{rf:Watanabe}. 
We observed all Raman active phonons of CuRh$_{2}$S$_{4}$ except for the T$_{2g}$ phonon with the lowest energy as shown in Fig. 4.
By compression, the position of the phonon peak monotonically shifts to the higher energy, suggesting simple hardening of the lattice. 
There is neither appearance of a new phonon peak nor indication of peak splitting.
Therefore, the micro-Raman spectroscopy reveals that there is no structural change across the MI transition.   
\begin{figure}
\includegraphics[height=72mm,clip]{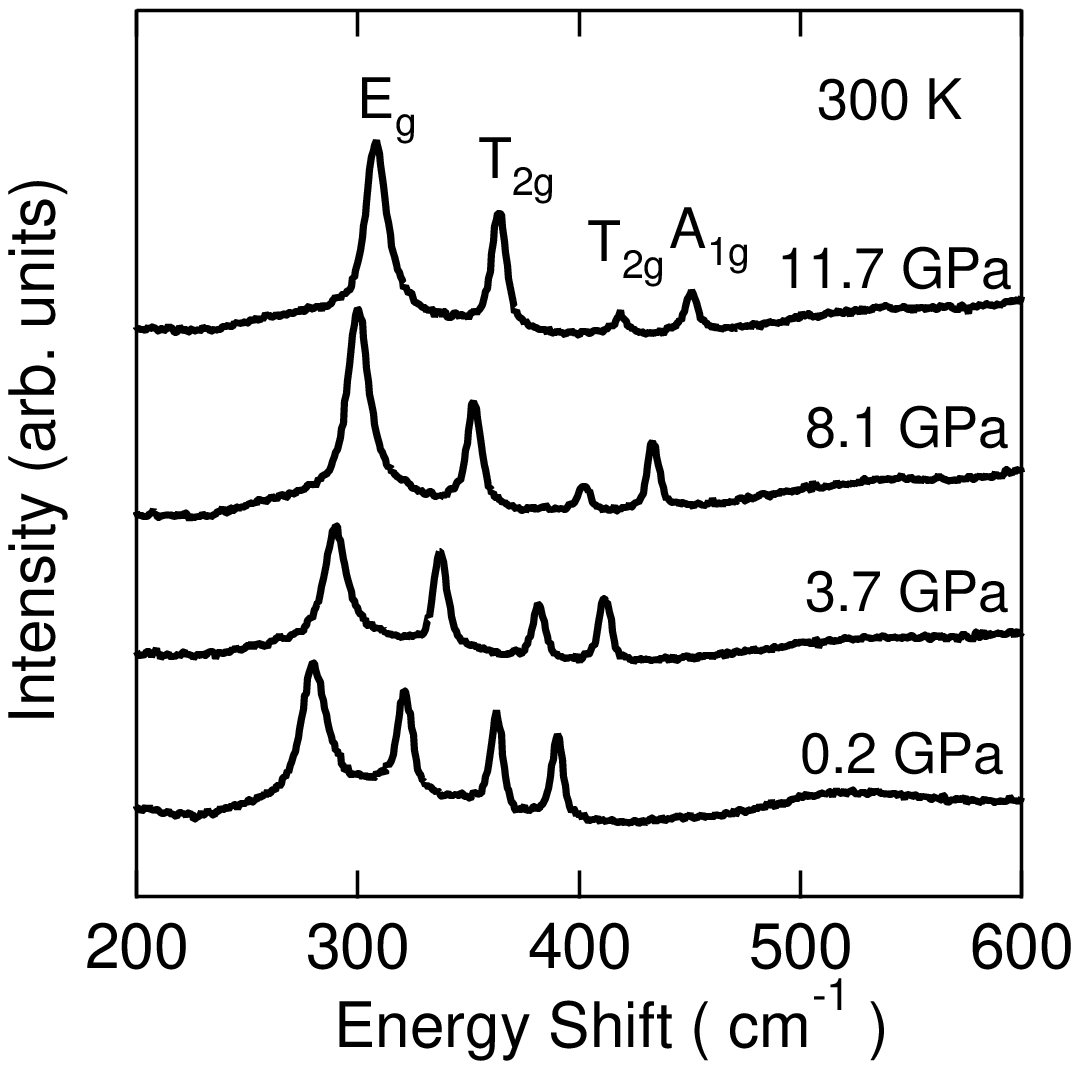}
\caption{\label{fig:fig.4} Pressure dependence of Raman spectra at 300 K. Each spectrum is shifted vertically to avoid the overlaps. 
There is neither appearance of a new phonon peak nor indication of peak splitting.
The relative change in the intensity among the peaks is arbitrary since the diameter of laser beam ($\sim 2\ \mu $m) is almost the same as the grain size of the single-crystalline CuRh$_{2}$S$_{4}$ and the position of the focused single-crystal in the sample possibly changes very slightly by compression.
}
\end{figure}

Another possible origin, which is not accompanied by a lattice change, is a modification of the band structure by pressurization, as reported for a divalent fcc-Yb crystal~\cite{rf:McWhan}. 
The fcc-Yb is known to be a material in which compression induces a semimetal-insulator transition without change of lattice symmetry. 
The transition is ascribed to the formation of an energy gap between two bands, both of which originally cross the Fermi level in the semimetallic phase~\cite{rf:Sankar}. 
According to the band calculation by Oguchi~\cite{rf:Oguchi}, the metallic conduction in CuRh$_{2}$S$_{4}$ is ascribable to the two holes in two bands, each of which crosses the Fermi level as the two bands of the fcc-Yb. $\rho(300\ \rm K)$ of CuRh$_{2}$S$_{4}$ increases gradually from 4.6$\times$10$^{-4}$ to 1.3$\times$10$^{-3}$Ħcm with compression of up to 8.0 GPa. 
This $P$-dependence is quite similar to that of fcc-Yb in which $\rho(300\ \rm K)$ also increases gradually from 3$\times$10$^{-5}$ Ħcm at ambient pressure to 3$\times$10$^{-4}$ Ħcm at 3.7 GPa.
Although CuRh$_{2}$S$_{4}$ is a good metal, an MI transition of the kind observed in the semimetal fcc-Yb, may be expected if the two bands near the Fermi level are sensitive to pressure. 
This scenario, based on the modification of the band structure, is likely to explain the pressure dependence of $T^{\ast}$ as well as the decrease of $T_{\rm c}$ prior to the SI transition.
The pressure-induced MI transition in non-superconducting CuIr$_{2}$Se$_{4}$~\cite{rf:Furubayashi_Se} may be understood by the same scenario.

In summary, 
we have found a pressure induced SI transition in a chalcogenide-spinel CuRh$_{2}$S$_{4}$ from resistivity measurements under quasi-hydrostatic pressure of up to 8.0 GPa. As pressure increases, the $T_{\rm c}$ value initially increases from 4.7 to 6.4 K and starts to slightly decrease after a broad maximum is reached.
With further compression, superconductivity vanishes suddenly at $P_{\rm SI}$ around 5.3 GPa. The SI transition occurs in concurrence with a metal-insulator transition induced by pressure. Judging from the variation of $\rho (300$ K$)$ and Raman spectra with pressure, a modification of the band structure is a possible origin for the pressure induced SI transition in CuRh$_{2}$S$_{4}$, as is the case with the pressure induced semimetal-insulator transition reported for an fcc-Yb crystal.
%
%
%

We thank T. Oguchi, P. Alireza and S. R. Julian for valuable discussions. This work was partially supported by a Grant-in-Aid for COE
Research (No. 13CE2002) and a Scientific Research (B) (No. 13440114) from the Ministry of Education, Culture, Sports, Science and Technology of Japan.
%
%
%
 
\end{document}